# Component based platform for multimedia applications

Ovidiu Rățoi[1], Haller Piroska[1], Ioan Salomie[2], Genge Bela[1]

[1]"Petru Maior" University of Tg. Mureș, N. Iorga st, No. 1, Romania, 540088
[2]Technical University of Cluj Napoca, Daicoviciu st, No. 15, Romania, 400020
[1]oratoi@engineering.upm.ro, {phaller, bgenge}@upm.ro
[2]Ioan.Salomie@cs.utcluj.ro

## Abstract

*We propose a platform for distributed multimedia applications which simplifies the development process and at the same time ensures application portability, flexibility and performance. The platform is implemented using the Netscape Portable Runtime (NSPR) and the Cross-Platform Component Object Model (XPCOM).*

## 1. Introduction

Recent years have shown an increased interest towards multimedia rich applications. Multimedia content ranges from text or simple images to audio data, video data or animations. The increased availability of broadband Internet connections leads the way towards applications that offer high quality multimedia streaming over wide area networks. In this context there is a need for solutions that enable application developers to quickly and effortlessly develop this kind of applications.

In the same time software components technologies and component based software engineering are maturing, in fact the use of components is a sign of maturity in any field of engineering. The usage of software components offers a lot of advantages the most important of them being reusability, a component once developed may be reused in any number of applications, depending of how generic are the services it offers. Also the task of applications developers changes from development of new software to composition of existing pieces. Another characteristic property of software components is encapsulation, the internal structure is hidden and it's services are accessed through a well defined interface. This property confers high flexibility to component based software as individual components can be easily replaced with improved ones as long as their interface remains identical.

Research and work in the area of multimedia content is not new, some years ago a standardization effort existed. In 1998 SC24, a subcommittee of ISO/IEC JTC 1 published PREMO (Presentation Environment for Multimedia Objects) [1] a standard for multimedia objects presentation and in the same time a reference model for distributed multimedia applications. They didn't propose new formats or standards for describing multimedia content instead they used the existing ones to offer a unified environment for application programming.

Network communication was always an important issue when handling multimedia content especially when real time streaming was involved. A research team [2] tackled this problem and proposed a multimedia applications middleware which regulated network traffic, optimized resource usage and offered a high degree of portability to applications.

Recently some groups are proposing a platform for collaboration systems which integrates mobile devices [3]. It uses a client-server architecture to provide multimedia content adapted to the capabilities of any devices used clients.

There were other attempts to use software components, in the context of multimedia applications, [4] proposes an architecture for real time video applications based on CORBA. The focus was on the quality of service by monitoring the system entities load and network communication. In time the use of general purpose components was proved not to be well suited for multimedia based systems, and dedicated frameworks were developed [5]. More recently [6] proposes a component based solution for multimedia stream synchronization.

The platform proposed in this paper is based on the notion of *channels*, denoting a communication link between two remote parties. The novelty of our approach is that the platform supports not only binary streaming (through binary channels), but also a collection of channels communicating through the use of messages specific to *web services*.

The paper is structured as follows. In section 2 we provide a detailed description of the proposed platform. In section 3 we describe the interface exposed by the platform and used by applications, and in section 4 we describe the interface exposed by the channel and used by the platform. We end with a conclusion and future work in section 5.

## 2. A Platform for Distributed Multimedia Applications

A well known method for providing a high degree of transparency and portability to distributed applications is positioning an intermediate layer called Multimedia Platform between the operating system and the application. In *Figure 1* the multilayer structure of a multimedia applications platform is presented.

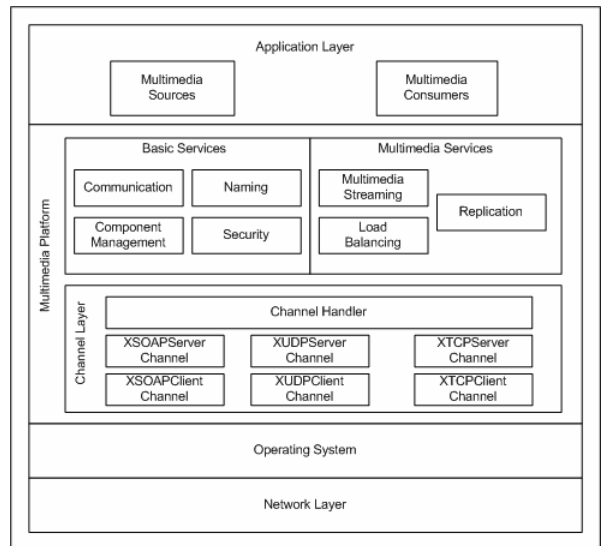

*Fig. 1. – Multimedia Platform*

The bottom layer in this architecture is represented by the network which provides host computer interconnection and basic data transmission services. On the following level we have the operating system which provides services ranging from process management or memory organization to communication and synchronization. Above the operating system we can find the Multimedia Platform which is divided into two sections: one channel layer (described later in this paper) and a service layer. The last layer of this architecture is represented by the application layer which contains multimedia sources or multimedia consumers.

The Multimedia platform offers a service for data streaming between multimedia sources and consumers. Whenever a consumer needs data from a source a stream between them has to be established. The communication is based on the concept of *channels*.

The Multimedia platform provides an interface for using the channels, named *XChannelHandler*. By using this interface we can create or destroy a channel, send messages to one specific channel and get the received messages from all channels. Each newly created channel is given a unique channel ID. All operations involving channels are done through this ID.

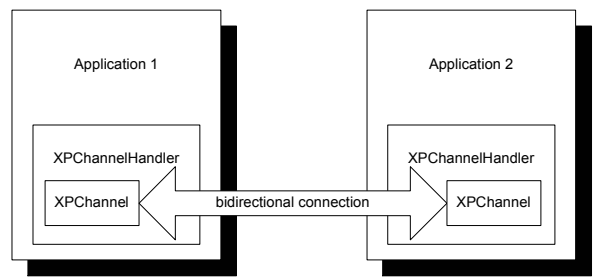

*Figure 2. Connection Model Architecture*

A *channel*, as mentioned in the previous section, is a logical communication link between two end-points, like presented in *Figure 2*. The style of peer to peer communication requires the existence of a connection between each pair of communicating applications. Channels embody communication protocols, while access is provided through one single interface.

In the current form the platform provides six types of channels, but the Multimedia platform architecture offers an easy way to add more channels.

For the moment applications can chose from the following channel types: *XSOAPServerChannel* (channel accepting SOAP client connections), *XSOAPClientChannel* (SOAP client channel), *XUDPServerChannel* (channel accepting UDP client connections), *XUDPClientChannel* (UDP client channel), *XTCPServerChannel* (channel accepting TCP client connections) and *XTCPClientChannel* (TCP client channel). Each type of channel represents one component, which is loaded by the platform. The interface implemented by the channels is the same for all of them.

The reason for introducing the concept of the UDP server and client channels was to create a similarity between the concept of a TCP connection and a UDP one. By using these channel types, applications that make use of TCP channels can be very easily switched to UPD channels, without having to change any of the application's upper-layers, or the logic on which the application was built. When a datagram is received using UDP channels, the source IP and port are analyzed. If they are not present in the saved internal list of the *XUDPServerChannel* channel, this means it is a "new connection" and a corresponding notification message is constructed for the upper-layers. Also, the newly received IP-port pair is saved in the internal list and a new channel is created which is provided with a new channel ID.

The SOAP channels are created for developing applications based on WEB Services architecture. It provides SOAP communication protocol for creating both server and client applications. The main problem with today's distributed systems is interoperability. Web Services intend to solve this problem by introducing software components that are "capable of being accessed via standard network protocols such as but not limited to SOAP over HTTP" [5].

SOAP [6] (i.e. Simple Object Access Protocol) provides a simple mechanism for exchanging structured and typed information through the form of XML messages. In order for our platform to support a standard web service interface, we created a SOAP-based channel capable of sending and receiving standard SOAP messages. For the implementation of the SOAP transport we used the open source *gSOAP* [7] library.

The SOAP standard does not only provide a means for exchanging XML data, but also binary data through the use of base64 or hex encodings. Because of this, integrating streaming data into SOAP messages becomes a straight-forward process.

SOAP messages are encoded through the form of envelopes, containing namespace definitions and a SOAP body. The body contains the actual messages. Our platform implements a *rawDataMessage* in the SOAP body for exchanging streaming binary data. An example SOAP message constructed by our channel is the following:

```
<?xml version="1.0" encoding="UTF-8"?>
<SOAP-ENV:Envelope
xmlns:SOAP-ENV=
"http://schemas.xmlsoap.org/soap/envelope/"
 xmlns:SOAP-ENC=
"http://schemas.xmlsoap.org/soap/encoding/"
 xmlns:xsi= http://www.w3.org/1999/XMLSchema-instance
```

```
xmlns:xsd= "http://www.w3.org/1999/XMLSchema"
xmlns:ns= "urn:simple-calc">
<SOAP-ENV:Body
SOAP-ENV:encodingStyle=
"http://schemas.xmlsoap.org/soap/encoding/">
<ns:rawDataMessage>
<data xsi:type= "xsd:base64Binary">
QUxBIEJBTEEgUE9SVE9DQUxBÂ□Â□
</data></ns:rawDataMessage>
</SOAP-ENV:Body>
</SOAP-ENV:Envelope>
```

## 3. Platform interfaces description

As we mentioned earlier the Multimedia platform includes support for SOAP, TCP and UDP channels, each of them implemented through the form of a component.

For accessing and managing those channels the Multimedia platform provides one interface named: *XChannelHandler*.

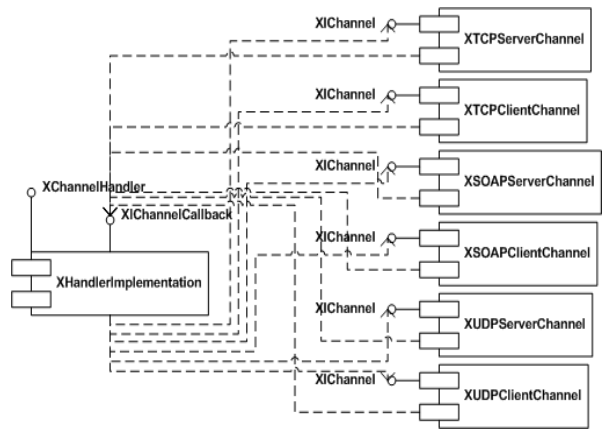

*Figure 3. General view of the channel architecture*

The *XChannelHandler* interface exposes four functions used for creating and destroying channels, sending messages to one channel or receiving messages from the opened channels.

- *int createChannel(channel_info*, unsigned int&)*. This function is used for creating one new channel. The *channel_info* is a structure which contains information for the channel which has to be created: channel type, host address and host port. Each time a channel is created a new channel *ID* is generated. This value is returned through the second argument of the function. The function returns CHANNEL_OK on success or a negative error code otherwise.

- *int destroyChannel( const unsigned int )*. This function is used for destroying one channel, identified by the channel *ID*. The function returns CHANNEL_OK on success or a negative error code otherwise.

- *int sendToChannel( const XMessage* )*. This function is used for sending one message to a channel. The message is encapsulated into an *XMessage* object which also contains the channel ID. The function returns CHANNEL_OK on success or a negative error code otherwise.

- *int getMessage( XMessage* & )*. This function is used for retrieving one message form the internal queue. All the messages received from the active channels are deployed into this internal queue. The source channel ID is contained into the message object. The function returns CHANNEL_OK if a message has been received or CHANNEL_NOMESSAGES otherwise.

## 4. Channels interfaces description

For interconnectivity issues all the communication components extend the same interface: *XIChannel*. This interface has the following functions:

- *int createChannel( void )*, which is called when a new channel has to be created. The function must return CHANNEL_OK on success, or a negative number denoting an error code.

- *int destroyChannel( void )*, which is called when a existing channel has to be destroyed. The function must return CHANNEL_OK on success, or a negative number denoting an error code.

- *int getChannelStatus( void )*, which is called for interrogating the state of the channel (i.e. connection). If the connection is OK, the function must return the CHANNEL_OK value. In case of error it returns CHANNEL_SOCKETERR or another negative value denoting an error code.

- *void addMessage( const XMessage*)*, which is used for adding one message to the outgoing queue of the channel. This function has to be a non blocking one.

All the received incoming messages received by the channels have to be added to one queue in the platform's architecture. This is done by using the interface named *XIChannelCallback*, which provides a single function:

- *int onChannelMessage( const XMessage*)*. This is a non blocking function, which provides a way for incoming messages to be added into a single queue in the platform architecture.

## 5. Conclusions and future work

In this paper we proposed a component based platform for distributed multimedia applications. This approach simplifies the development of multimedia centered applications and ensures their transparency, portability and performance. By providing a unique interface for all supported channel types, application developers can easily change the underlying transport and channel type.

As future work, we intend to extend the supported channel collection with security by adding a new layer on existing UDP, TCP and SOAP channels. We also intend to implement additional web service technology support for UDDI, WSDL and the semantic web through SAWSDL and OWL-S.